\documentclass[global,twocolumn]{svjour}
\usepackage[T1]{fontenc}
\usepackage{times}
\usepackage[ngerman,british]{babel}
\usepackage[amssymb]{SIunits}
\usepackage{graphicx}
\usepackage{amsmath,amsfonts,amssymb,dsfont}
\usepackage{mathcomp}

\begin{document}
\title{An ultraviolet laser system for laser cooling and trapping of metastable magnesium}
\author{A. P. Kulosa\and J. Friebe\and M. P. Riedmann\and A. Pape\and T. W. W\"ubbena\and D. B. Fim\and S. R\"uhmann\and K. H. Zipfel\and H. Kelkar\and W. Ertmer\and and E. M. Rasel}
\institute{Leibniz Universit\"at Hannover, Welfengarten 1, D-30167 Hannover, Germany\\phone: +49 (0)511-762-19203, fax: +49 (0)511-762-2211, \email{rasel@iqo.uni-hannover.de}}

\maketitle						

\begin{abstract}
We report on a reliable laser system for cooling magnesium atoms in the metastable $^{3}P$ manifold.  The three relevant transitions coupling the $^{3}P$ to the $^{3}D$ manifold are near 383~nm and seperated by several hundred $\giga\hertz$. The laser system consists of three diode lasers at $766\,\nano\meter$. All lasers are frequency stabilised to a single pre-stabilised transfer cavity. The applied scheme for frequency control greatly reduces the complexity of operating three lasers combined with resonant frequency doubling stages and provides a high reliability necessary for complex atomic physics experiments.
\end{abstract}

\section{Introduction}
Laser manipulation of atoms is an indispensable tool in atom optics. Most experiments require the laser frequency to be stabilised, typically to an atomic transition. This can be achieved via frequency modulation spectroscopy in an atomic vapour cell if the transition originates from the ground state \cite{galzerano}. Applying this method to metastable transitions involves additional complexity of an electric discharge and the associated linewidth broadening \cite{hansen}. In such cases, as well as in the case that the laser frequency is far from any atomic transition, one can transfer the stability of a reference laser to the target laser via a transfer cavity or a frequency comb \cite{riedle,bohlouli,liekhus}. The transfer cavity technique has been demonstrated for metastable transitions in strontium where the lasers are stabilised to independent reference cavities \cite{xu}. This approach considerably increases the complexity of the system with the number of lasers involved. Locking multiple lasers to a single cavity has been achieved using a fast scanning cavity offset lock \cite{seymour}. However, the system has a feedback bandwidth limited by the scan frequency of the cavity.

\begin{figure}[tp]
	\centering
		\includegraphics[width=0.35\textwidth]{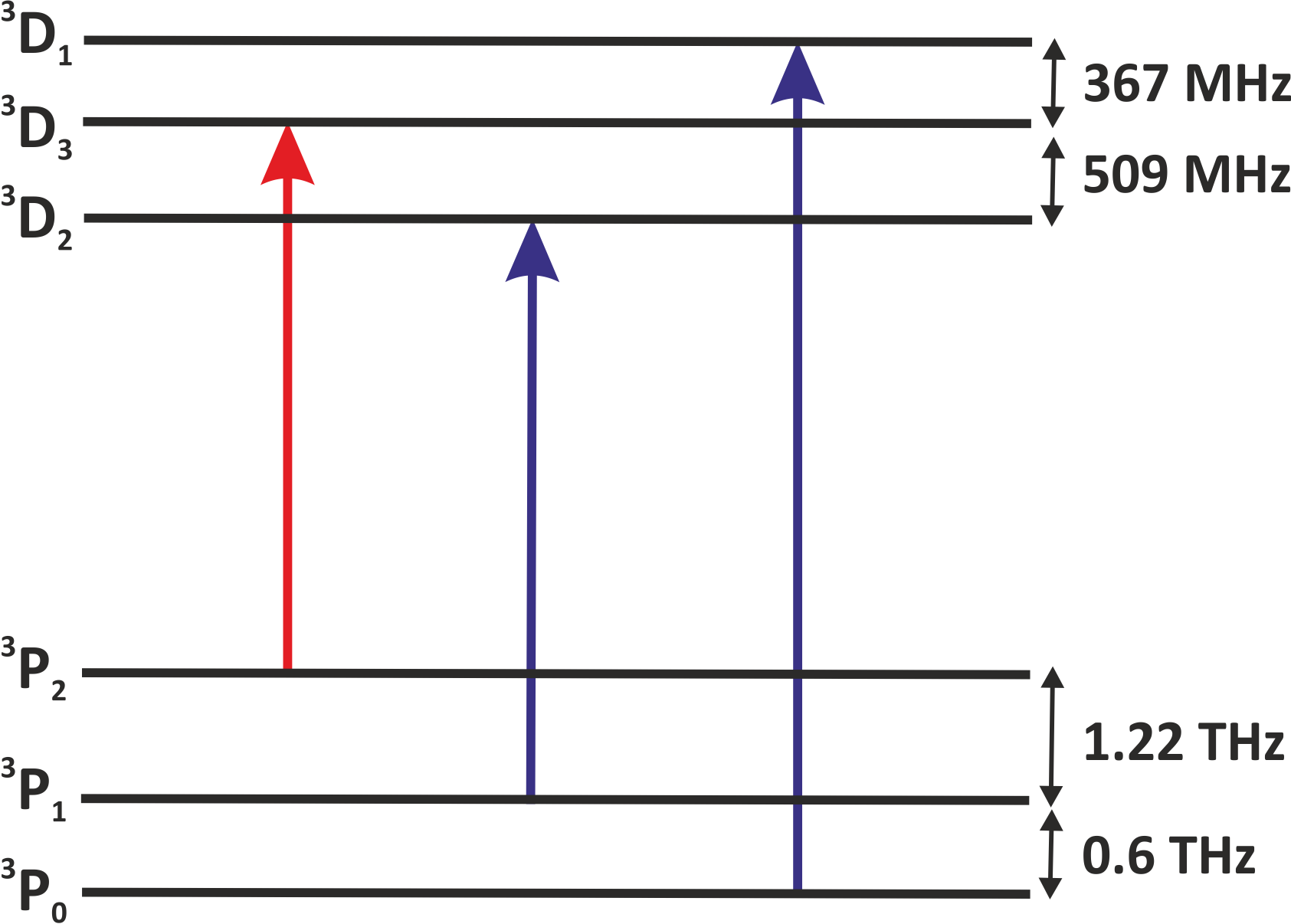}
	\caption{Relevant energy level scheme of the triplet manifold of $^{24}\text{Mg}$. The red arrow depicts the closed cooling transition, whereas the blue arrows represent the repumping transitions.}
	\label{fig:Levelscheme}
\end{figure}

\begin{figure}[htp]
	\centering
		\includegraphics[width=0.5\textwidth]{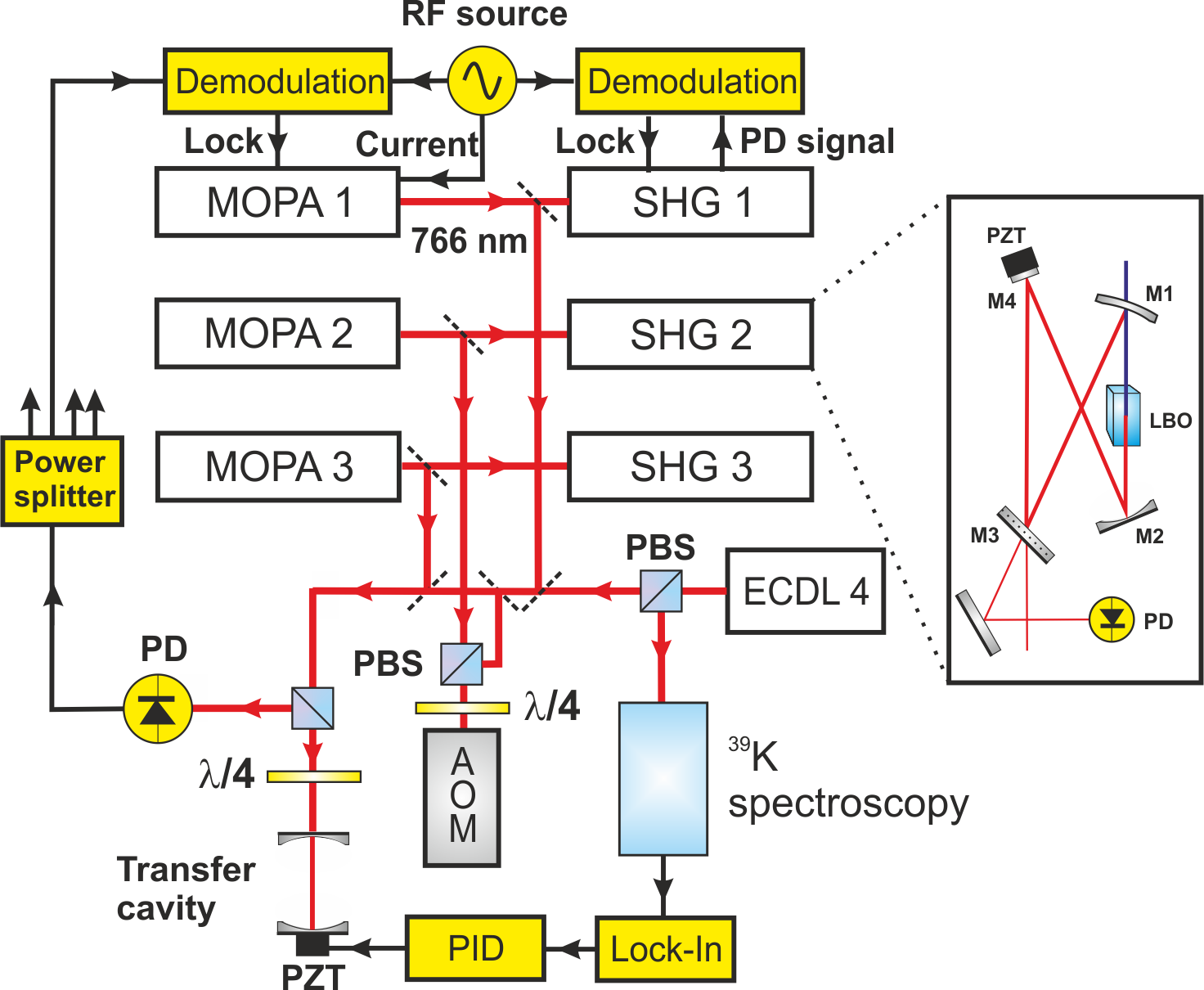}
	\caption{Simultaneous locking of three independent laser systems to a common transfer cavity referenced to an atomic transition. Each laser system is locked to the cavity using the PDH scheme. The required modulation is generated by RF local oscillators imprinting sidebands on the laser frequency via diode current modulation. The same modulation is also used to stabilise the individual SHG cavity of each MOPA. The modulation frequencies are seperated by a few MHz to allow the unperturbed demodulation of each signal. The length of the transfer cavity is stabilised with the help of a fourth laser referenced to the $\text{D}_{2}$-line of $^{39}\text{K}$ via Doppler-free spectroscopy. The locking scheme is shown in detail only for MOPA 1 for clarity. Blow-up: Bowtie cavity design for second harmonic generation. (PZT: piezo-electric transducer, PBS: polarising beam splitter, PD: photo diode)} 
	\label{fig:Lasersystem}
\end{figure}

In this article, we demonstrate the transfer of frequency stability of the reference laser to three independent lasers using a single transfer cavity. These lasers are around $766\,\nano\meter$ and are frequency doubled to obtain $383\,\nano\meter$ light needed to manipulate magnesium atoms in the metastable $^{3}P$ manifold. The relevant metastable energy levels of magnesium are shown in Fig. \ref{fig:Levelscheme}. Each of the three laser systems consists of a master oscillator power amplifier (MOPA) feeding a cavity enhanced second harmonic generation (SHG) stage. The MOPA systems are locked to the reference cavity that is pre-stabilised to the $\text{D}_{2}$-line of $^{39}\text{K}$ at $766\,\nano\meter$. We apply the Pound-Drever-Hall (PDH) stabilisation scheme based on diode current modulation \cite{drever}. The modulation sidebands are also used to lock the SHG cavities to the light at $766\,\nano\meter$. A schematic setup of the complete laser system is shown in Fig. \ref{fig:Lasersystem}. 

\section{The MOPA system}
The three seed lasers are external-cavity diode lasers (ECDL) in Littrow configuration \cite{ricci} with an output power of $50\,\milli\watt$. The laser diodes are anti-reflection (AR) coated and emit light at a central wavelength of 770$\,\nano\meter$ with a tuning range of $\pm40\,\nano\meter$. The external cavity consists of a grating with 1800~lines/mm. According to the Bragg condition for the first order $\lambda=2d\cdot\sin(\alpha)$, this results in a lasing operation for $\lambda=766\,\nano\meter$ at an incident angle $\alpha\approx 43.6\,^{\circ}$.
The temperature of the laser diodes is stabilised to few $\milli\kelvin$ and an optical isolator ($60\,$dB) protects the laser from back reflections. A tapered amplifier with a central wavelength of $765\,\nano\meter$ is used for power amplification which is followed by an isolator ($35\,$dB) to suppress back reflections. Each MOPA system has an output power of $1\,\watt$. A Voigt function fit to the beat measurement performed between two free-running MOPAs with a sweep time of $530\,\milli\second$ and a resolution bandwidth (RBW) of $10\,\kilo\hertz$ resulted in Gaussian and Lorentzian linewidths (FWHM) of $260\,\kilo\hertz$ and $35\,\kilo\hertz$, respectively.

The frequency of each MOPA system is stabilised by locking the master lasers to a transfer cavity via the PDH method. MOPA 2, in addition, passes through an acousto-optic modulator (AOM) in double-pass configuration before the transfer cavity which allows the frequency to be scanned around the $^{3}P_{2}\rightarrow\,^{3}D_{3}$ cooling transition.

The frequency modulation sidebands for the PDH lock are generated by current modulation around $20\,\mega\hertz$. The differences in modulation frequency are kept large enough to distinguish the error signals of the individual MOPA. Each error signal is fed to the corresponding laser diode current for fast frequency control and to a piezo-electric transducer (PZT) for high gain at low Fourier frequencies.

\section{The length stabilised transfer cavity} 
The transfer cavity consists of a plane and a concave mirror (with radius of curvature $R=100\,\milli\meter$) for $766\,\nano\meter$ seperated by a quartz glass spacer of length $L=7\,\centi\meter$. One of the mirrors is attached to a PZT for controlling the length of the cavity. The finesse is determined to be $\mathcal{F}\approx200$ while the linewidth of the resonances is $10.2\,\mega\hertz$. In order to lock the laser to an atomic transition frequency, it is necessary that the cavity resonance be within a few 100 MHz of the transition frequency. This can be achieved by generating a higher order mode degeneracy creating a spectrum of equidistant resonances with a spectral separation $\Delta\nu=c/(2NL)$, where $N\in\mathds{N}$. Following the approach in \cite{budker}, this can be accomplished by choosing a specific length of the cavity according to the relation

\begin{equation}\frac{L}{R}=1-\cos^2{\left(\frac{q\pi}{N}\right)}.
\label{equation1}
\end{equation}

Here $q\in\mathds{N}$, $q<N$ and $q$, $N$ are mutually prime. In the experiment, $q=6$ and $N=19$, resulting in a mode spacing of $112.5\,\mega\hertz$. 

\begin{figure}[tp]
	\centering
		\includegraphics[width=0.5\textwidth]{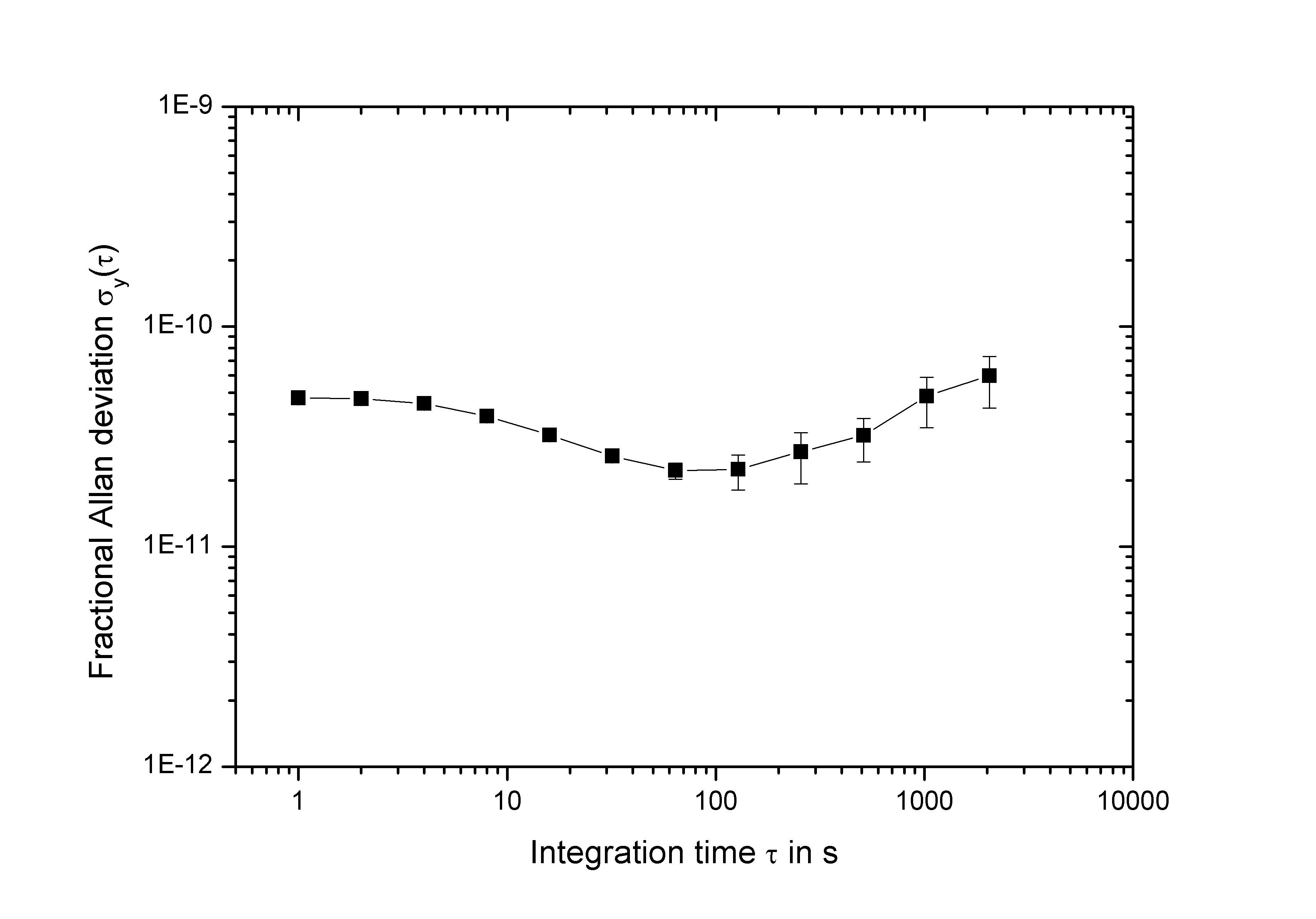}
	\caption{Fractional frequency instability of the beat note between MOPA 1 and MOPA 2 that are seperated by $200\,\mega\hertz$ and locked to the transfer cavity. }
	\label{fig:adev}
\end{figure}

The absolute length of the cavity is stabilised to an ECDL with an output power of $P=45\,\milli\watt$ locked to the $^{39}\text{K}$ $\text{D}_{2}$-line at $766\,\nano\meter$ via Doppler-free saturation spectroscopy. The cavity is heated above room temperature to make fine adjustments to the cavity length to satisfy Eqn. \ref{equation1} and also for temperature stabilisation. 

The locking performance of the transfer cavity was determined in the time domain by locking MOPA 1 and MOPA 2 to adjacent modes of the cavity with a servo bandwidth of a few $100\,\kilo\hertz$. The laser frequencies were seperated by $200\,\mega\hertz$. The beat signal between the two lasers was detected with a photo diode and fed to a frequency counter with zero dead-time. The gate time for the measurement was $1\,\second$. From the time series the Allan deviation was calculated. The result is shown in Fig. \ref{fig:adev}. The relative frequency instability reaches a value of $\sigma_{y}\approx 2\times 10^{-11}$ for integration times $\tau\approx 100\,\second$. The flicker floor of the lasers results from a jitter in the frequency control loop. 

\section{The second harmonic generation}  
\begin{figure}[tp]
	\centering
		\includegraphics[width=0.50\textwidth]{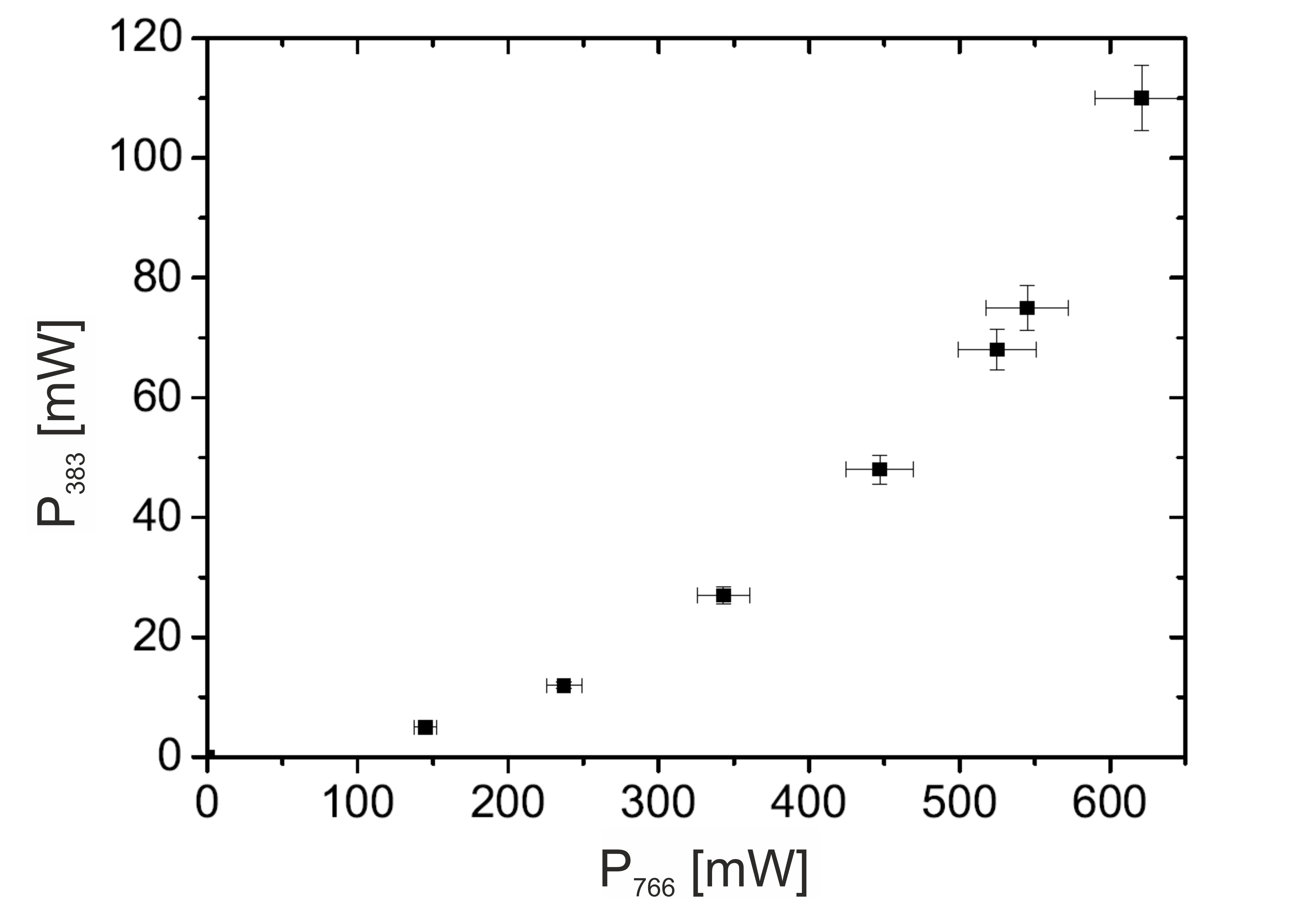}
	\caption{Generated UV power at $383\,\nano\meter$ as a function of the power of the incident light field at $766\,\nano\meter$.}
	\label{fig:UVpower}
\end{figure}

Second harmonic generation in a $15\,\milli\meter$ long LBO crystal placed inside a bow-tie cavity delivers more than $100\,\milli\watt$ of power at $383\,\nano\meter$, corresponding to a conversion efficiency of $\approx 20\,\%$. Fig.~\ref{fig:UVpower} shows the generated power in the UV as a function of the incident power at $766\,\nano\meter$. The crystal is AR coated on both sides. The outcoupling mirror M1 is a minescus (radius of curvature $R=\pm50\,\milli\meter$) to minimise the output beam divergence. Mirror M1 together with curved mirror M2 (radius of curvature  = $50\,\milli\meter$), focus the beam to a waist of $30\,\micro\meter$ inside the crystal. The incoupling plane mirror M3 has a transmission of $1.2\,\%$. Mirror M4 is $4\,\milli\meter$ in diameter and attached to a PZT (Fig. \ref{fig:Lasersystem}). The total length of the cavity is about $28\,\centi\meter$. The finesse of the cavity is calculated to be $\mathcal{F}=270$ from the measured linewidth of $\delta\nu=3.93\pm0.2\,\mega\hertz$. The crystal is temperature stabilised to room temperature.

The SHG cavity is stabilised to the light at $766\,\nano\meter$ with the PDH scheme using the modulation sidebands generated by diode current modulation. The lock has a bandwidth of $2.5\,\kilo\hertz$. The power fluctuations in the output beam are below $2\,\%$. A beat measurement performed between two lasers at $383\,\nano\meter$ resulted in a Gaussian linewidth of $1.2\,\mega\hertz$ and a Lorentzian linewidth of $25\,\kilo\hertz$ measured with a RBW of $100\,\kilo\hertz$ and a sweep time of $4\,\milli\second$. 

The laser system is used to realise a magneto-optical trap (MOT) for metastable $^{3}P_{2}$ magnesium atoms loaded from a MOT operated on the $^{1}S_{0}\rightarrow\,^{1}P_{1}$ transition in the singlet manifold. $10^9$ atoms at $3\,\milli\kelvin$ are transferred to the triplet manifold using the narrow intercombination transition ($^{1}S_{0}\rightarrow\,^{3}P_{1}$) at $457\,\nano\meter$. $1.6\,\text{x}\,10^8$ atoms are then captured by the "triplet-MOT" at $383\,\nano\meter$ where they are further cooled to $1\,\milli\kelvin$. This serves as starting point for loading them into a dipole trap or a magic wavelength lattice for creating an optical lattice clock.
                                                                                                                                             
\section{Summary}
We established frequency stabilisation of three lasers at $766\,\nano\meter$ to a single reference cavity whose length is pre-stabilised using a fourth laser locked to a $^{39}\text{K}$ spectroscopy signal. The light of the target lasers is frequency doubled to obtain more than $100\,\milli\watt$ of UV light at $383\,\nano\meter$. Sidebands generated by laser diode current modulation are used for locking the lasers to the transfer cavity as well as for locking the second harmonic generation cavities to the fundamental light. Our setup reduces the complexity of operating multiple frequency stabilised lasers. In principle, it allows in the present configuration to lock up to nine lasers to the cavity limited only by its finesse. The method can also be applied to other experiments requiring multiple stabilised lasers by an appropriately designed cavity. A few examples are the sub-Doppler cooling of calcium in the metastable manifold \cite{hansen}, the strontium optical clock \cite{xu} and the simultaneous magneto-optical trapping of $^{6}\text{Li}$, $^{40}\text{K}$ and $^{87}\text{Rb}$ \cite{taglieber}, among many others.

\end{document}